\documentclass[reprint,twocolumn,superscriptaddress,amsmath,amssymb,aps]{revtex4-1}

\usepackage{graphicx}
\usepackage{graphics}
\usepackage{amsmath}
\usepackage{dcolumn}
\usepackage{bm}

\begin{document}


\title{Electronic structure, magnetism, and optical properties of orthorhombic GdFeO$_3$\\ from first principles}

\author{Xu-Hui Zhu}
\affiliation{Beijing National Laboratory for Condensed Matter Physics, Institute of Physics, Chinese Academy of Sciences, Beijing 100190, China}
\affiliation{Institute of Atomicand Molecular Physics, College of Physical Science and Technology, Sichuan University, Chengdu 610065, China}
\author{Xiang-Bo Xiao}
\affiliation{Beijing National Laboratory for Condensed Matter Physics, Institute of Physics, Chinese Academy of Sciences, Beijing 100190, China}
\author{Xiang-Rong Chen}
\affiliation{Institute of Atomicand Molecular Physics, College of Physical Science and Technology, Sichuan University, Chengdu 610065, China}
\author{Bang-Gui Liu}\email{bgliu@iphy.ac.cn}
\affiliation{Beijing National Laboratory for Condensed Matter Physics, Institute of Physics, Chinese Academy of Sciences, Beijing 100190, China}
\affiliation{School of Physical Sciences, University of Chinese Academy of Sciences, Beijing 100190, China}

\date{\today}

\begin{abstract}
Orthorhombic GdFeO$_3$ has attracted considerable attention in recent years because its magnetic structure is similar to that in the well-known BiFeO$_3$ material. We investigate electronic structure, magnetism, and optical properties of the orthorhombic GdFeO$_3$ in terms of density-functional-theory calculations. The modified Becke-Johnson (mBJ) exchange potential is adopted to improve on the description of the electronic structure. Our calculation show that the G-type antiferromagnetic (G-AFM ordering of Fe spins) phase of orthorhombic GdFeO$_3$ is stable compared to other magnetic phases. The semiconductor gap calculated with mBJ, substantially larger than that with GGA, is in good agreement with recent experimental values. Besides, we also investigate effect of the spin-orbit coupling on the electronic structure, and calculate with mBJ the complex dielectric functions and other optical functions of photon energy. The magnetic exchange interactions are also investigated, which gives a Neel temperature close to experimental observation. For comparison towards supporting our results, we study the electronic structure of rhombohedral (R3c) BiFeO$_3$ with mBJ. These lead to a satisfactory theoretical understanding of the electronic structure, magnetism, and optical properties of orthorhombic GdFeO$_3$ and can help elucidate electronic structures and optical properties of other similar materials.
\end{abstract}

\pacs{Valid PACS appear here}
\maketitle


\section{Introduction}

Bismuth ferrite (BiFeO$_3$) is the representative of single-phase multiferroic materials, which displays antiferromagnetic order below $T_N$$\sim$$643$ K and possesses relatively high spontaneous electric polarization of 59.4 $\mu$C/cm$^2$
until $T_c$$\sim$$1100$ K \cite{1,2,3}. Besides, BiFeO$_3$ is a perovskite whose most stable phase is a rhombohedral distorted structure with space group R3c. As a magnetic materials similar to BiFeO$_3$, the orthorhombic distorted GdFeO$_3$, which has the Neel temperature  661 K and belongs to the perovskite rare-earth orthoferrites, has sparked substantially curiosities and stimulated relatively deeper research \cite{4,5,6,7}. 

The GdFeO$_3$ compound has a complex $H$-$T$ phase diagram and undergoes a plurality of magnetic phase transition, accompanying the dramatic changes in the electrical properties \cite{8}. The orthorhombic distorted GdFeO$_3$ compound (Pbnm), with Gd$^{3+}$ ions at the center and Fe$^{3+}$ ions at the corners surrounded by oxygen octahedra, possesses weak ferromagnetism and ferroelectricity \cite{4,5,7}. Under electric and magnetic fields, the ferroelectric polarization and magnetization of GdFeO$_3$ compound has been successfully brought under control for wide applications \cite{9}. A spontaneous polarization of about 0.12 $\mu$C/cm$^2$ was obtained at 2 K \cite{7}, which is basically identical with the measured value in the perpendicular magnetic system \cite{10}. According to the Bertaut notation, the spin structure of Fe$^{3+}$ is G$_x$A$_y$F$_z$ \cite{11}. Below $T^{\rm Gd}_N$=2.5 K, the magnetic order of Gd$^{3+}$ is antiferromagnetic along $a$-axis, showing G$_x$ antiferromagnetic order and ferroelectric polarization characteristic \cite{4,12}. The interaction between adjacent Fe$^{3+}$ and Gd$^{3+}$ layer induces the ferroelectric polarization along the $c$-axis \cite{12}. It was reported that the Fe spins in BiFeO$_3$ form a G-type antiferromagnetic (G-AFM) order, with the spins on the Fe$^{3+}$ ions being aligned anti-ferromagnetically along the [111] axis \cite{13}. Interestingly, GdFeO$_3$ can show a ferromagnetism below 5 K \cite{4,14,15}, and at $T^{\rm Fe}_N$=661 K, Fe$^{3+}$ also forms the G-AFM order in GdFeO$_3$ and shows a weak ferromagnetism due to the Dzyaloshinskii-Moriya interaction \cite{16,17,18}. On the other hand, the first-principles calculation suggested that the antiferromagnetic phase of the orthorhombic GdFeO$_3$ is more stable than the ferromagnetic phase \cite{19}, and it was also pointed out that the transition from antiferromagnetic to paramagnetic ordering occurs at 670 K \cite{7,6,9}.

Despite a large number of experimental studies concerning the electromagnetic of orthorhombic distorted GdFeO$_3$, the theoretical reports are extremely meager and it is necessary to use theoretical approaches to perform further study. Here, we investigate the electronic structure and magnetic and optical properties of the orthorhombic GdFeO$_3$ through density functional theory (DFT) calculation. In order to better understand the electronic properties of GdFeO$_3$, we also investigate the electronic structure of BiFeO$_3$ for comparison. The rest of this paper is organized as follows. We shall describe our computational details in the second section. We shall present our main calculated results and analysis in the third section. Finally, we shall give our conclusion in the fourth section.

\section{Computational details}

The full-potential linearized augmented plane wave method within the density-functional theory (DFT) \cite{20,21}, as implemented in the package Wien2k \cite{22}, is utilized in our calculation. Firstly, the popular generalized gradient approximation (GGA-PBE) \cite{23} is adopted to optimize crystal structures and investigate electronic structures and magnetism. Because the standard semi-local GGA usually underestimates energy band gaps \cite{24}, we use the mBJ approximation \cite{25} for the exchange potential, taking the local density approximation (LDA) \cite{26} to treat the correlation potential, as usual for improved description of electronic structures and optical properties. For electronic structure calculations, the mBJ has been demonstrated to significantly improve and produce accurate semiconductor gaps for sp semiconductors, wide-band-gap semiconductors, and transition-metal oxide semiconductors and insulators \cite{25,27,28,29,30}. Because the theoretical semiconductor gaps are improved, much better computational results can also be obtained for the optical properties. The full relativistic effects are calculated with the Dirac equations for core states, and the scalar relativistic approximation is used for valence states \cite{31,32}. We also take the spin-orbit coupling (SOC) into consideration. The cut-off energy is set to -6 Ry to separate core states from valance states. The k-mesh size in the first Brillouin zone is 11$\times$10$\times$7 for GdFeO$_3$, and 10$\times$10$\times$10  for BiFeO$_3$. We make harmonic expansion up to $l_{\rm max}$=10, set $R_{\rm mt}$$\times$$K_{\rm max}$=7, and use magnitude of the largest vector $G_{\rm max}$=12 in charge density Fourier expansion. The radii of Gd, Bi, Fe, and O atomic spheres are set to 2.27, 2.12, 1.99, and 1.77 bohr, respectively. The self-consistent calculations are considered to be converged only when the integration of absolute charge-density difference per formula unit between the successive loops is less than $0.0001|e|$, where $e$ is the electron charge.

\section{Results and discussion}

\begin{figure}[!htbp]
{\centering  
\includegraphics[clip, width=8cm]{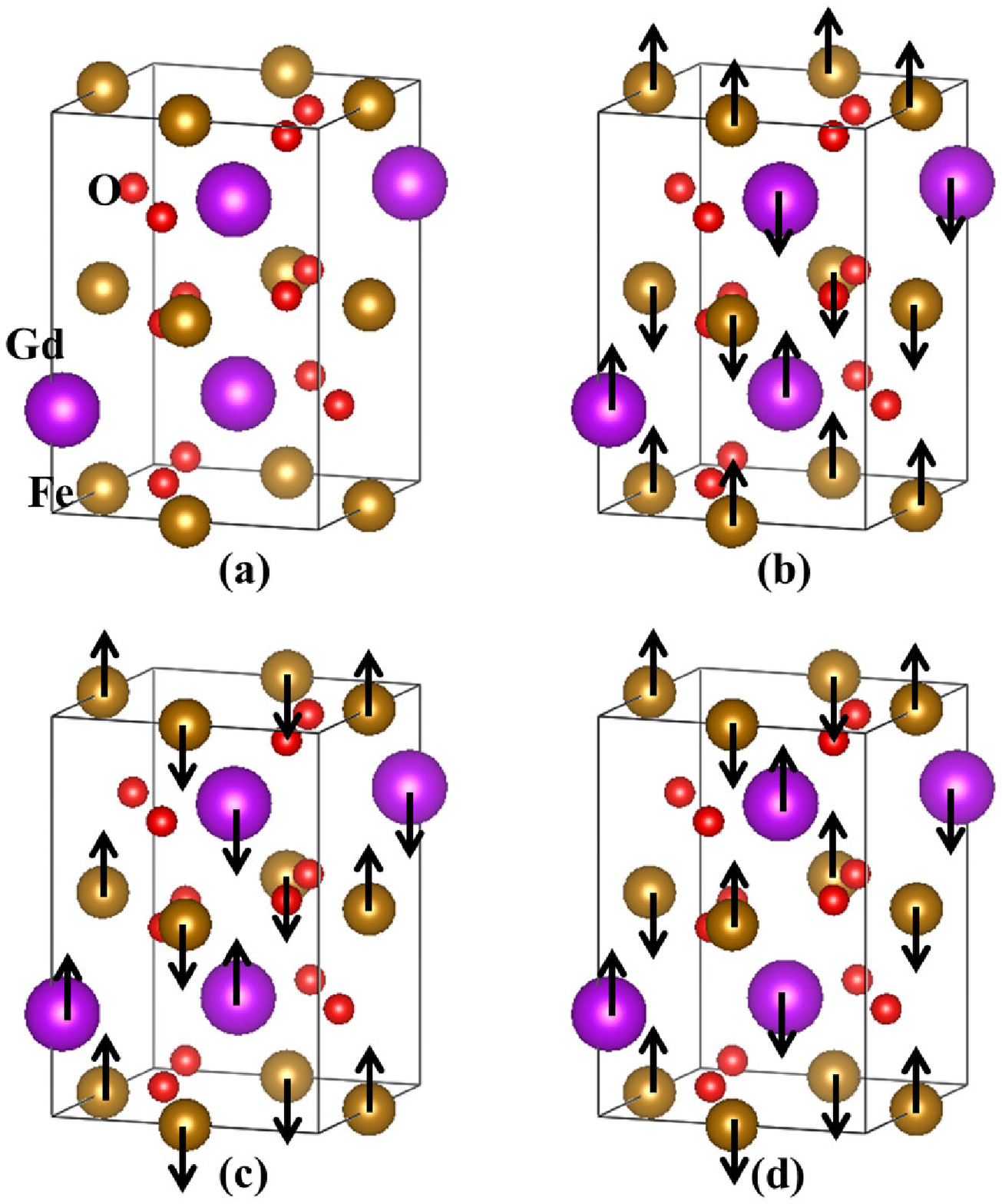}}\\
\caption{(Color online) The crystal structure and three antiferromagnetic ordering configurations in orthorhombic GdFeO$_3$: (b) A-AFM, (c) C-AFM, and (d) G-AFM. The arrows indicate magnetic moment orientations on Gd and Fe atoms.}\label{fig1}
\end{figure}

\subsection{Crystal structure}

The orthorhombic GdFeO$_3$ investigated here has space group Pbnm (No.62) \cite{4,19,33,34} at low temperatures. The experimental lattice constants are $a = 5.349$ \AA, $b = 5.611$ \AA, and $c = 7.669$ \AA{} \cite{33}. We demonstrate the crystal structure of the orthorhombic GdFeO$_3$ in Fig. 1(a). At first, we optimize the lattice parameters and ionic positions with GGA and LDA. The optimized lattice parameters are summarized in Table I. Existing experimental results \cite{33,34} are also presented for comparison. The GGA optimized volume $V$ is 2.0\% larger than the experimental volume \cite{33}, but the LDA optimized volume $V$ is 5.1\% smaller. It can be clearly seen that the lattice constants and volume calculated with GGA are closer to the experimental data \cite{33,34} than those with LDA. Therefore, the other properties are investigated on the basis of the GGA optimized result. The GGA-optimized atomic positions are summarized in Table II. After internal structure optimization, the Gd atom occupies the ($0.9911, 0.0639, 0.25$) site, the Fe atom the ($0, 0.5, 0$) site, the O1 atom  the ($0.7024, 0.3121, 0.0488$) site, and the O2 atom the ($0.0927, 0.4726, 0.25$) site in Wyckoff coordinates. They are consistent with the experimental orthorhombic structure \cite{33,34}. Furthermore, the three optimized Gd-O bond lengths are 2.307 \AA, 2.347 \AA, and 2.399 \AA, slightly larger than Fe-O bond lengths of 1.968 \AA, 1.974 \AA, and 2.124 \AA{}. These are in line with the relation of ionic radii, Gd$^{3+}$ $>$ Fe$^{3+}$, and in accordance with the previous reported values \cite{34}. The distances between Gd$^{3+}$ and Fe$^{3+}$ are 3.136 \AA, 3.284 \AA, and 3.362 \AA. The bond angles of Fe-O-Gd are 85.38$^\circ$ and 87.34$^\circ$, deviating from the ideal values of 90$^\circ$. These imply that the orthorhombic GdFeO$_3$ has undergone great structural distortion.

\begin{table}[htp]
\caption{The lattice parameters optimized with LDA and GGA and experimental data of the GdFeO$_3$.}
\begin{ruledtabular}
\begin{tabular}{cccccc}
	& $a$(\AA) &	$b$(\AA)	& $c$(\AA)	& $V$(\AA$^3$)	& $\alpha=\beta=\gamma$ ($^{\circ}$)\\ \hline
GGA	& 5.399	 & 5.714	& 7.612	    & 234.83	        &90.0\\
LDA	& 5.222	 & 5.620	& 7.446	    & 218.51	& 90.0\\
Exp.\cite{33}	& 5.349	 & 5.611	& 7.669	& 230.17	& 90.0\\
Exp.\cite{34}	& 5.351	 & 5.612	& 7.671	& 230.38	& 90.0
\end{tabular}
\end{ruledtabular}
\end{table}

\begin{table}[htp]
\caption{The atomic positions ($x,y,z$) optimized with GGA of the GdFeO$_3$, in comparison with experimental values.}
\begin{ruledtabular}
\begin{tabular}{ccccc}
Atom	& site	& $x$ (exp.[33,34])	& $y$ (exp. \cite{33,34})	& $z$ (exp. \cite{33,34}) \\ \hline
Gd	& 4c	& 0.9911  	& 0.0639 	& 0.2500 \\
    &       & ($0.9844, 0.9846$) 	& ($0.0628, 0.0626$)	& ($0.2500, 0.2500$)\\ \hline
Fe	& 4b	& 0.0000	& 0.5000  	& 0.0000 \\
    &       & ($0.0000, 0.0000$)	& ($0.5000, 0.5000$)  	& ($0.0000, 0.0000$)\\ \hline
O1	& 8d	& 0.7024 	& 0.3121 	& 0.0488\\
    &       & ($0.6957, 0.6966$)	& ($0.3016, 0.3011$)	& ($0.0506, 0.0518$)\\ \hline
O2	& 4c	& 0.0927 	& 0.4726 	& 0.2500 \\
    &       & ($0.1005, 0.1009$)	& ($0.4672, 0.4669$)	& ($0.2500, 0.2500$)
\end{tabular}
\end{ruledtabular}
\end{table}

In addition, we have calculated the GGA total energies of four different magnetic ordering configurations: ferromagnetic and three antiferromagnetic (AFM) ones. The three AFM structures are shown in Fig. 1(b-d) and denoted by A-AFM, C-AFM, and G-AFM, respectively. Taking the total energy of the lowest G-AFM structure as a reference, the other three energies are higher, which is consistent with the experimental results \cite{4,8}. This ground-state magnetic structure is similar to the magnetic ordering of rhombohedral BiFeO$_3$ where the Fe spins form a G-AFM structure \cite{35}. These results are also consistent with LSDA+U calculation \cite{19}.

\subsection{Electronic structures}

With the optimized crystal structure, we calculate with both GGA and mBJ potentials the spin-dependent energy band structure and the densities of states (DOSs) of the orthorhombic GdFeO$_3$ between -6 eV and 4 eV. The two band structures are shown in Fig. 2. The conduction band bottom and the valence band top are located at the same S point in both of the band structures. This means a direct gap for the orthorhombic GdFeO$_3$. It can be seen that the GGA semiconductor gap is 0.61 [Fig. 2(a)] eV, a little larger than earlier first-principles result 0.54 eV \cite{36}, but the mBJ-calculated semiconductor gap, 2.49 eV [Fig. 2(b)], is apparently larger than the GGA value and is in accordance with the experimental results \cite{7,9,10}. Very interestingly, there are many similar features between GdFeO$_3$ and BiFeO$_3$ as  orthoferrite ABO$_3$ materials. As a typical multiferroic orthoferrites compound, however, BiFeO$_3$ shows the characteristics of indirect band gap. For the BiFeO$_3$, our GGA band gap of 0.965 eV is slightly lower than previous DFT values of 1.06 eV \cite{38} and 1.04 eV \cite{39}, but they are all too small to describe the experimental values of 2.4 eV \cite{40}, 2.5 eV \cite{41}, and 2.74 eV \cite{42}. Our mBJ calculation produces a semiconductor gap of 2.354 eV for BiFeO$_3$ and it is in good agreement with the experimental values. In contrast, a band gap of 2.8 eV, obtained with screened exchange potential \cite{35}, is too large to describe the experimental values. These show that our mBJ gap of 2.49 eV is reasonable and should be accurate  for the orthorhombic GdFeO$_3$.

\begin{figure}[!tbp]
\centering  
\includegraphics[clip, width=8.5cm]{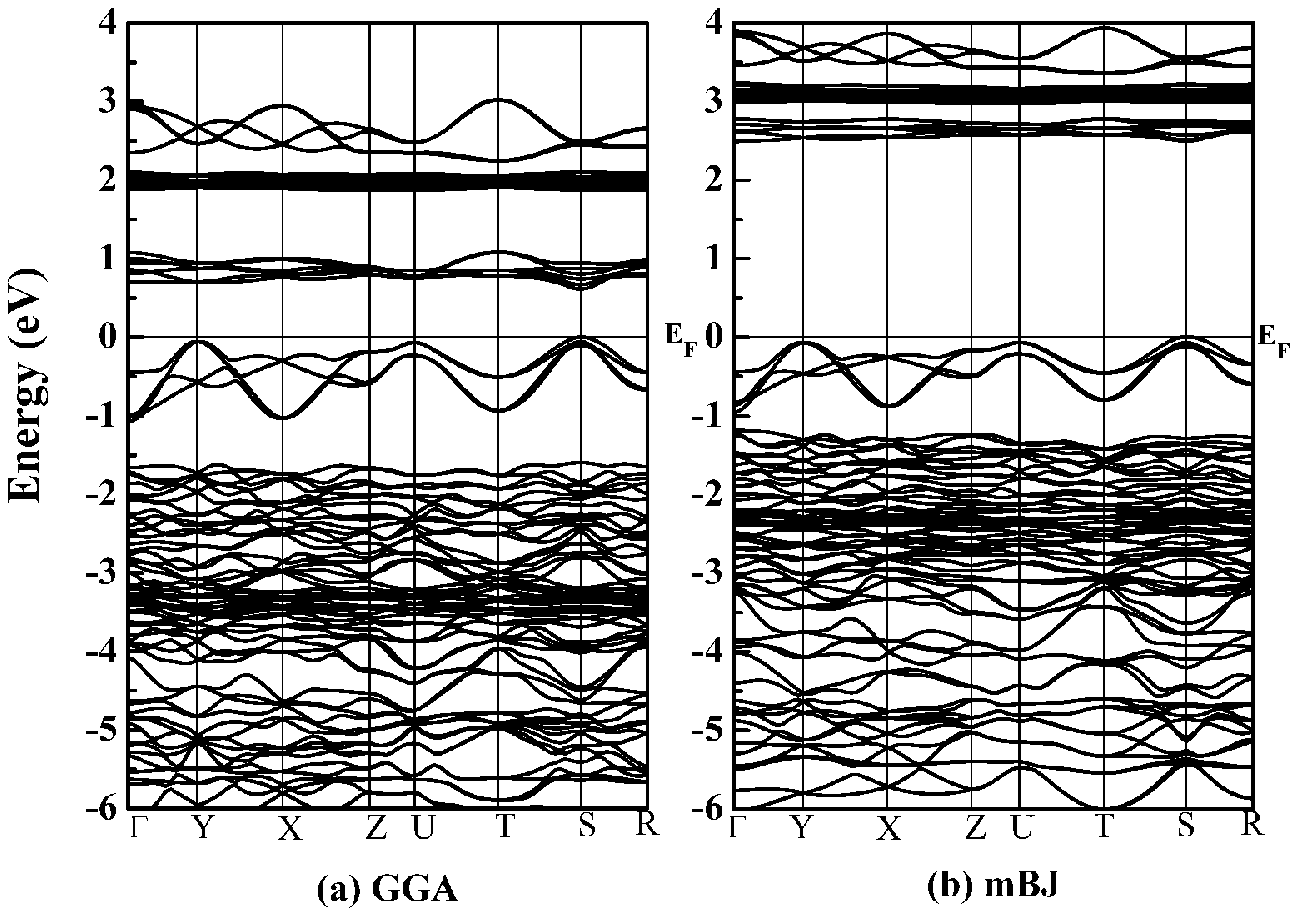}\\
\caption{(Color online) The spin-resolved energy bands of the orthorhombic GdFeO$_3$ with (a) GGA and (b) mBJ.}\label{fig2}
\end{figure}

In Fig. 3 we present the spin-resolved densities of states (DOSs) of the orthorhombic GdFeO$_3$ calculated with both GGA and mBJ. Through comparing GGA DOS [see Fig. 3(a)] and mBJ DOS [see Fig. 3(b)], we can see that the wide valence bands between -6 eV and 0 eV are originated from O 2p and Fe 3p states with a mixture of some Gd 5p6s, and the conduction bands are mainly from Fe 3d and Gd 4f states. Our analysis shows that the filled O 2p states are located between -6 and 0 eV. The filled Gd 4f states are between -3.4 and -1.2 eV, and the empty ones are between 2.8 and 3.4 eV. The filled Fe eg states are between -0.85 and 0.0 eV, and the empty ones between 2.49 and 2.93 eV. The empty Fe t2g states are between 3.37 and 3.88 eV. In order to understand the electronic properties of the orthorhombic GdFeO$_3$, the total and partial density of states of BiFeO$_3$ are also investigated with mBJ (not presented here). The top of valence bands consist mainly of O 2p states and some Fe 3d and Bi 6p states, and the bottom of conduction bands are originated from Fe 3d states and  O 2p states. Our DOS calculated with mBJ is significantly more accurate than the previous theoretical work with GGA \cite{38,39}, and however is close to the sX potential result \cite{35}. It is interesting that the two materials share main features in the density of states.

\begin{figure}[!tbp]
\centering  
\includegraphics[clip, width=7.5cm]{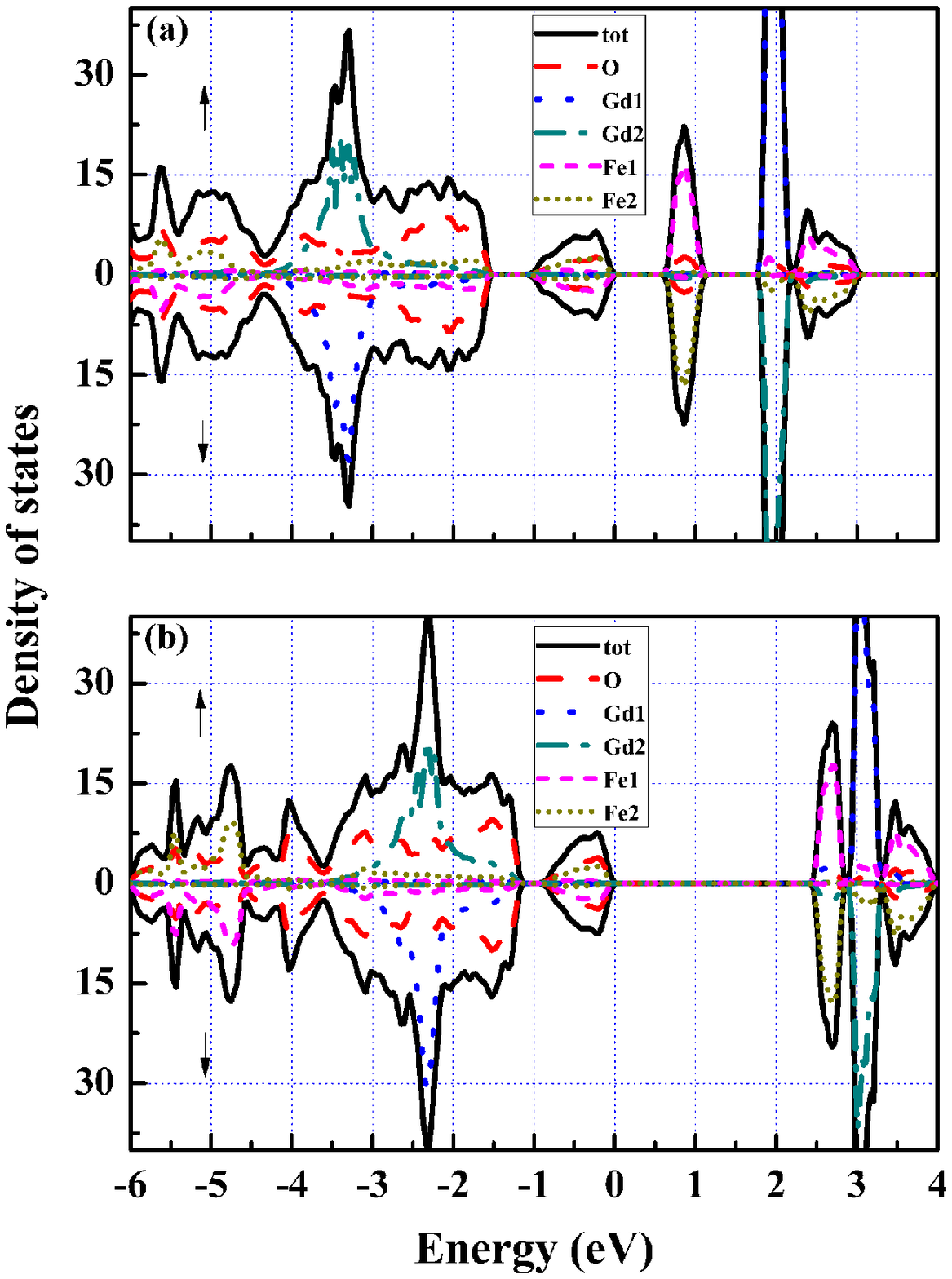}\\
\caption{(Color online) Spin-resolved total density of states of the orthorhombic GdFeO$_3$ with (a) GGA and (b) mBJ. The upper part is for majority-spin channel and the lower for minority-spin one.}\label{edge1}
\end{figure}

\subsection{Effects of the spin-orbits coupling}

\begin{figure}[!tbp]
\centering  
\includegraphics[clip, width=8.5cm]{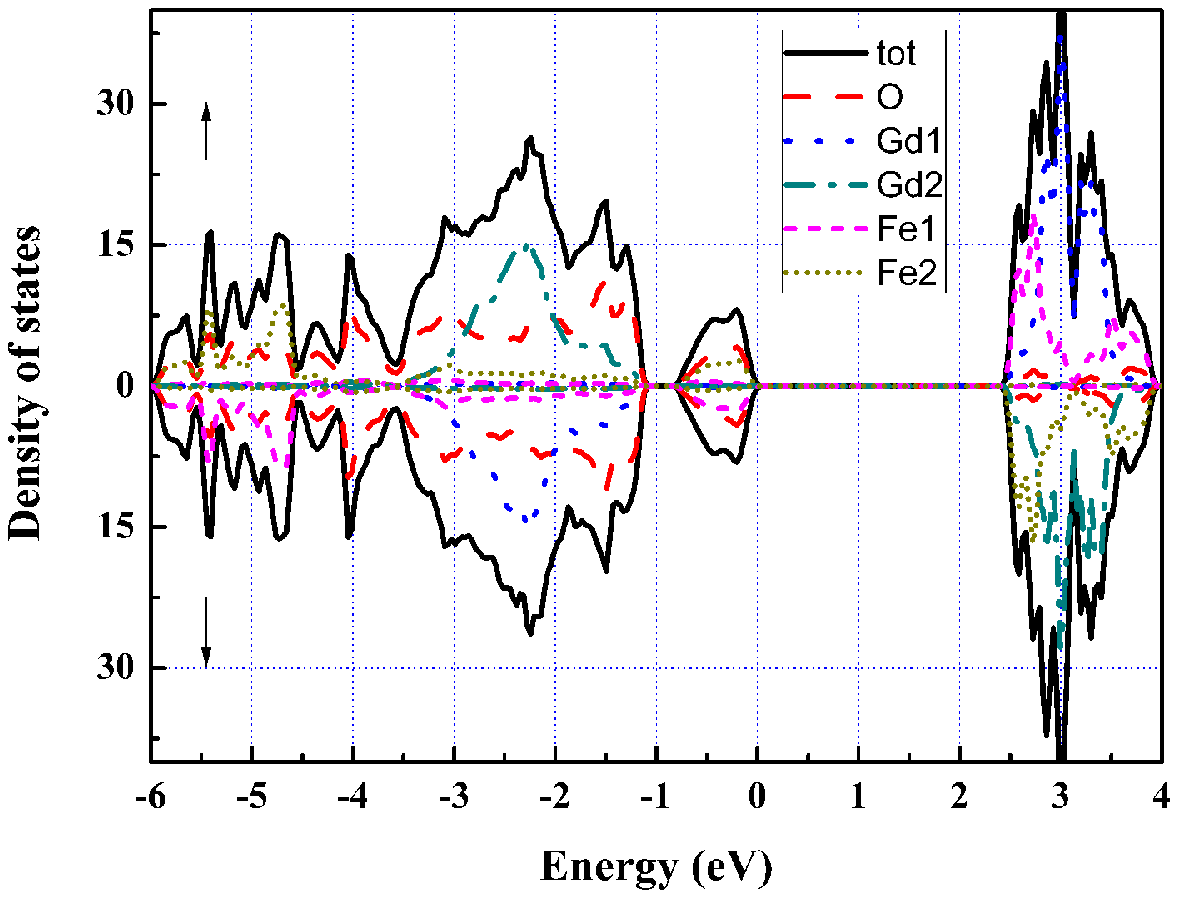}\\
\includegraphics[clip, width=8.5cm]{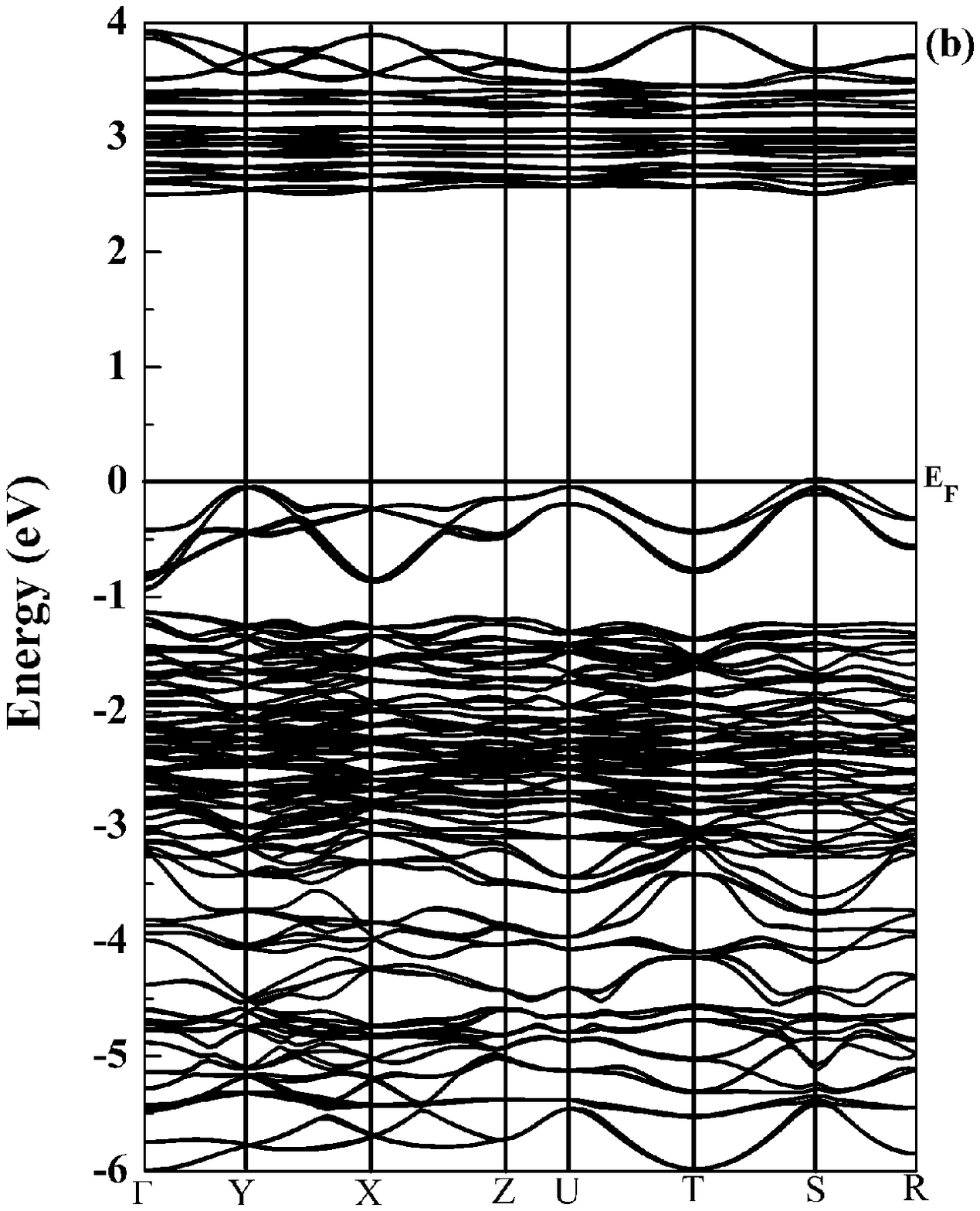}
\caption{(Color online) The Spin-resolved (a) total density of states and (b) energy bands of orthorhombic GdFeO$_3$ with mBJ+SOC.}\label{edge2}
\end{figure}

The spin-orbit coupling (SOC) is important to electronic materials including heavy atoms such as Gd. It can cause magnetocrystalline anisotropy. With GGA+SOC method, we calculate total energy of the orthorhombic GdFeO$_3$ by taking the SOC into account. Setting the magnetization in the [100], [010], [001], [110], [101], [011], and [111] directions, we  obtain the total energies: 2.6, 3.8, 0, 3.2, 8.7, 1.3, and 1.6 $\mu$eV, respectively. It is obvious that the lowest energy is along the [001] direction. These indicate that the easy magnetization axis (the most stable magnetic orientation) of the orthorhombic GdFeO$_3$ is along the [001] axis.

In the easy axis, the total spin moment is precisely equivalent to $0 \mu_B$ per formula unit without SOC for the orthorhombic GdFeO$_3$. According to the Hund's rule, the cations Gd$^{3+}$ and Fe$^{3+}$ possess the high spin values of $s = 7/2$ and $s = 5/2$, respectively, and the antiferromagnetic coupling makes the total spin moment equal to $0 \mu_B$ per formula unit. Since part of the spin moments are located in the interstitial region, the spin moments of the individual Gd$^{3+}$ and Fe$^{3+}$ are $6.855 \mu_B$ and $4.082 \mu_B$, smaller than the theoretical $7 \mu_B$ and $5 \mu_B$, respectively. When taking SOC into account, the spin moments of Gd$^{3+}$ and Fe$^{3+}$ reduce to $6.811 \mu_B$ and $4.080 \mu_B$, respectively. The orbital moment of Fe 3d is $0.183 \mu_B$, which has the same sign as the spin moment, and the orbital moment of Gd$^{3+}$ ion is $0.088 \mu_B$.

The semiconductor gap $E_g$ is also investigated by using mBJ. The semiconductor gap becomes smaller 2.40 eV when SOC is taken into account. We present in Fig. 4(a) the density of states of the orthorhombic GdFeO$_3$ obtained with mBJ+SOC. Looking closely at the Fig. 4(a), the semiconductor gap is slightly smaller than that without SOC. This should be because the Fe 3d and Gd 4f bands  become wider due to SOC. Fig. 4(b) explicitly demonstrates the energy bands with mBJ+SOC. The band structures and density of states show that  the energy bands, especially the conduction bands, in both of the spin channels hybridize with each other.

\subsection{Exchange interactions}

The magnetic exchange interactions on the Gd$^{3+}$ and Fe$^{3+}$ can be investigated in terms of total energy calculations. We consider four different magnetic configurations, namely the antiferromagnetic (the ground state) and the ferromagnetic order, and two other magnetic orders constructed by changing the Fe and Gd spins of the antiferromagnetic configuration to the ferromagnetic order, respectively.
With the first principles total energies of the different magnetic configurations, we can determine the coupling constants of the effective Heisenberg model \cite{43} $H=\sum_{ij}J_{ij}S_i\cdot S_j$, where $S_i$ is the spin operator at site $i$. Here, the summation is over spin pairs, and the spin exchange parameter $J_{ij}$ is limited to the nearest (Fe-Gd) and the next nearest (Fe-Fe and Gd-Gd) spin pairs. Although the magnetic moments in the spheres of Gd and Fe are $6.855 \mu_B$ and $4.082 \mu_B$, the Gd$^{3+}$ and Fe$^{3+}$ cations should theoretically contribute 7 $\mu_B$ and 5 $\mu_B$, respectively. We can assign spin values $s = 7/2$ and $s = 5/2$ to the Gd and Fe spins, respectively. Accordingly, there exists a relation between the magnetic energies $e_{ij}$ and exchange parameter $J_{ij}$, $e_{ij} = J_{ij}s_is_j$, where $s_i$ takes either 7/2 or 5/2. Taking the G-AFM ground state as a reference, the calculated total energies of other three states are 159.9 meV, 158.6 meV, and 3.5 meV per formula unit, respectively. The total energy can be split into $E_0 + \sum_{ij}e_{ij}$, where $E_0$ is defined to be independent of spin configuration. We obtain the following equations from the four magnetic structures.
\begin{equation}
\left\{ \begin{array}{l}
0=E_0-3e_{\rm Fe-Fe}-3e_{\rm Gd-Gd}\\
159.9=E_0+8e_{\rm Fe-Gd}+3e_{\rm Fe-Fe}+3e_{\rm Gd-Gd}\\
158.6=E_0+3e_{\rm Fe-Fe}-3e_{\rm Gd-Gd}\\
3.5=E_0-3e_{\rm Fe-Fe}+3e_{\rm Gd-Gd}
\end{array}\right.
\end{equation}
From the above equations, we can calculate the $e_{ij}$ parameters, and then obtain exchange coupling parameters $J_{ij}$. The Gd-Gd and Fe-Gd spin exchange energies  are much smaller than the Fe-Fe value 26.43 meV. As a result, the spin exchange parameters $J_{ij}$ are 0.03 meV between the nearest Fe-Gd pair, 4.23 meV between Fe-Fe, and 0.05 meV between Gd-Gd. It is clear that the Fe-Fe spin coupling is dominant over the other two. If neglecting the much smaller Gd-Gd and Gd-Fe interaction energies, we can estimate the Neel temperature, $T_N$=605 K, in terms of an analytical approach\cite{add43}. Considering that we have not take the SOC effect into account, this Neel temperature is very satisfactory compared to experimental value.

\begin{figure*}[!tbp]
\centering  
\includegraphics[clip, width=15cm]{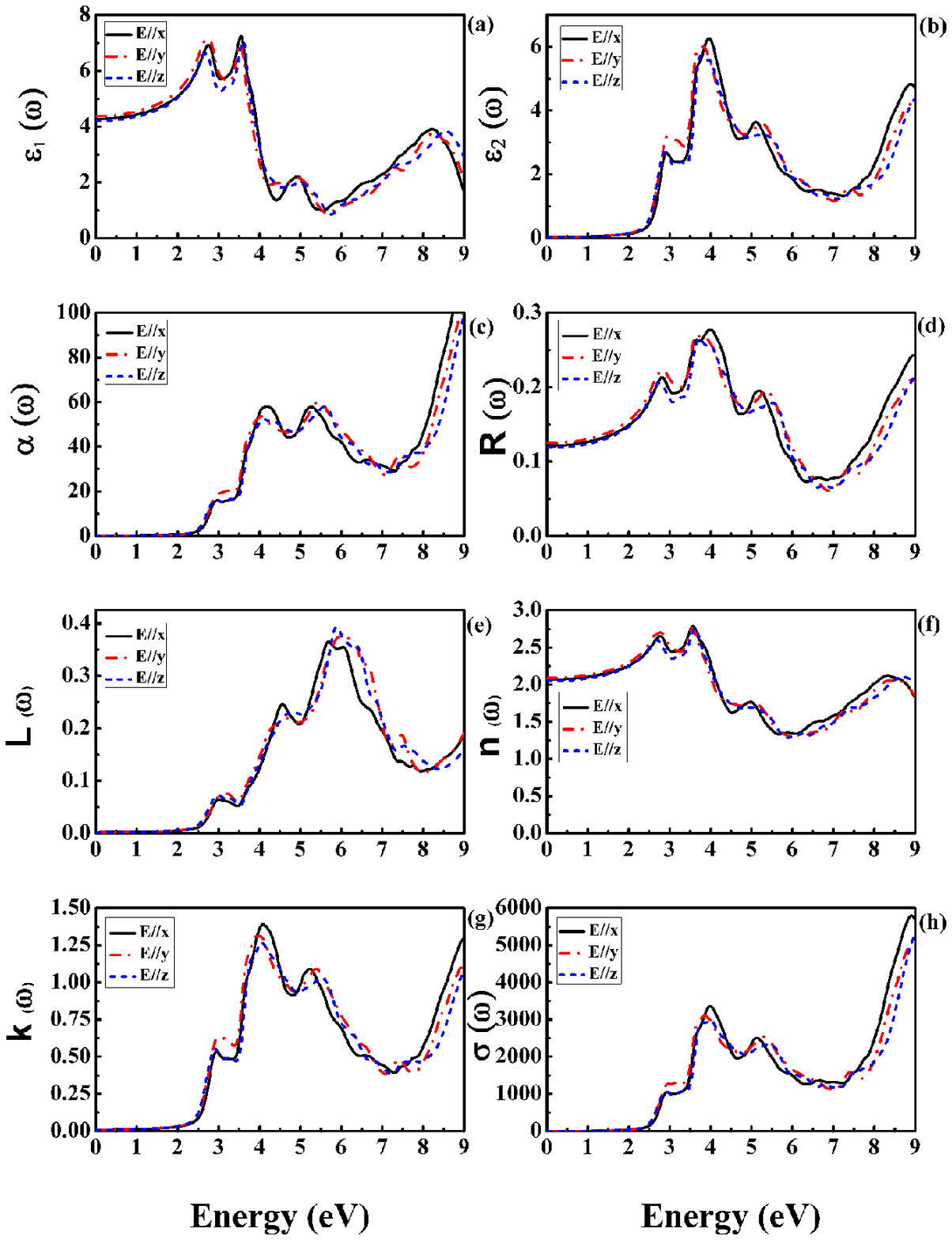}\\
\caption{Optical spectra as a function of photon energy for orthorhombic GdFeO$_3$ calculated with mBJ: (a) Real $\varepsilon_1(\omega)$, (b) imaginary $\varepsilon_2(\omega)$ parts of dielectric function, (c) absorption coefficient $\alpha(\omega)$, (d) reflectivity $R(\omega)$, (e) energy loss function $L(\omega)$, (f) refractive index $n(\omega)$, (g) extinction coefficient $k(\omega)$, and (e) optical conductivity $\sigma(\omega)$.}\label{edge}
\end{figure*}

\subsection{Optical properties}

The optical spectroscopy analysis is a powerful tool to determine the energy band structure of a solid material \cite{44,45}. The complex dielectric function is directly related to the energy band structure of solids. For the orthorhombic GdFeO$_3$, we present in Fig. 5 the mBJ calculated curves of the complex dielectric function (the real and imaginary parts), absorption coefficient, reflectivity, energy loss function, refractive index, extinction coefficient, and optical conductivity as functions of the photon energy in the range of 0-9 eV. All the three polarization directions ($E//x$, $y$, and $z$) are considered.

The electronic polarizability of a material can be understood from the real part of the dielectric function $\varepsilon_1(\omega)$ [Fig. 5(a)]. The static dielectric constant $\varepsilon_1(0)$ along the three crystallographic directions is found to be 4.28 for $E//x$, 4.38 for $E//y$, and 4.20 for $E//z$, respectively. The average value of zero frequency dielectric constant $\varepsilon_1(0)$ is 4.29. However, there is no experimental polarized zero frequency dielectric constant available for comparison. These results clearly indicate the anisotropy in the optical properties of orthorhombic GdFeO$_3$. The ratio $\varepsilon_1^{yy}(0)/\varepsilon_1^{zz}(0)$ is equal to 1.043 for estimating the degree of anisotropy. From zero frequency limit, they starts increasing and reaches the maximum value of 6.92 at 2.76 eV for $E//x$, 7.16 at 2.73 eV for $E//y$, and 6.63 at 2.65 eV for $E//z$, respectively. The imaginary part $\varepsilon_2(\omega)$ [Fig. 5(b)] gives the information of absorption behavior of the GdFeO$_3$. The threshold energy of the dielectric function is at $E_0 = 2.49$ eV, in accordance well with the fundamental gap. The obtained optical gap once again proves that mBJ can make accurate band gap for magnetic semiconductor. The imaginary part $\varepsilon_2(\omega)$ [Fig. 5(b)] indicates that the GdFeO$_3$ is anisotropic and its maximum absorption peak values are around 3.96, 3.82 and 3.77 eV for $E//x$, $E//y$ and $E//z$, respectively. From Fig. 4(a), for the imaginary part $\varepsilon_2(\omega)$, it is clear that there are strong absorption peaks in the energy range of 2.5-9 eV. Because the $\varepsilon_2(\omega)$ is related to the DOS, these peaks reflect some transitions between different orbitals. Compared with Fig. 3, it can be recognized that the peaks around 3.5-4.5 eV are mainly due to transitions from Gd-4f valence bands to O-2p conduction bands.

Fig. 5 (c)-(h) show the calculated results of the photon energy dependence of absorption coefficient $\alpha(\omega)$, reflectivity coefficient $R(\omega)$, energy loss function $L(\omega)$, refractive index $n(\omega)$, extinction coefficient $k(\omega)$ and optical conductivity $\sigma(\omega)$ of the orthorhombic GdFeO$_3$. The absorption coefficient $\alpha(\omega)$ [Fig. 5(c)] shows a very intense absorption up to 9 eV. It begins to increase sharply from 2.49 eV, corresponding to the band gap value. The reflectivity coefficient $R(\omega)$ is displayed in Fig. 5(d), the zero-frequency reflectivity are 12.1\% for $E//x$, 12.5\% for $E//y$, and 11.9\% for $E//z$, respectively. The maximum reflectivity values are about 19.5\%, 19.5\% and 17.9\%, which occurs at 5.18 eV for $E//x$, 5.35 eV for $E//y$, and 5.51 eV for $E//z$, respectively. Interestingly, the strong reflectivity maximum between 2.49 and 9 eV originates from the interband transitions. The energy loss function $L(\omega)$ [Fig. 5(e)] is related to the energy loss of a fast electron in the material and is usually large at the plasmon energy \cite{46}. The most prominent peak in $L(\omega)$ spectra represents the characteristic associated with the plasmon resonance and situates at 5.86 eV for $E//z$ polarization. The refractive index $n(\omega)$ are displayed in Fig. 5(f). The static refractive index $n(0)$ is found to have the value 2.07 for $E//x$, 2.09 for $E//y$, and 2.05 for $E//z$, respectively. The average value of $n(0)$ is equal to 2.07. The value of static refractive index is obtained from the real part of dielectric function to be $n(0) = \varepsilon_1(0)^{1/2} = \sqrt{4.29} = 2.07$, which is same as that obtained from Fig. 5(f). A similar trend is observed from the behaviour of the imaginary part of dielectric function $\varepsilon_2(\omega)$ [Fig. 5(b)] and the extinction coefficient $k(\omega)$ [Fig. 5(g)]. The extinction coefficient $k(\omega)$ reflects the maximum absorption in the medium at 4.01 eV for $E//x$, 3.90 eV for $E//y$, and 4.04 eV for $E//z$, respectively. The optical conductivity $\sigma(\omega)$ is shown in Fig. 5(h). It starts from 2.49 eV and have similar features with the absorption coefficient $\alpha(\omega)$ in Fig. 5(c).

\section{Conclusion}

We have used FP-LAPW method to investigate the structural, electronic, magnetic, and optical properties of orthorhombic GdFeO$_3$. The GGA approach has confirmed that the G-type AFM ordering of Fe spins is the ground-state phase, consistent with the experimental results. The mBJ exchange potential is used for improving on description of the electronic structures of the GdFeO$_3$. Our calculated results show that mBJ exchange greatly improves the accuracy of the band gap value. The mBJ result accords well with the experimental value and overcomes the GGA underestimation of the band gap. Besides, the spin-orbits coupling is taken into account to determine the easy magnetic axis and investigate its effect on the electronic structure. We also calculate magnetic exchange constants and thereby achieve a good Neel temperature close to the experimental value. Finally, the optical properties also are investigated with mBJ. In addition, we also calculate electronic structure of the well-known BiFeO$_3$ to support our calculated results in the case of the GdFeO$_3$. The magnetic similarity between these two perovskite oxide materials are very interesting. These calculated results should be useful to obtain more insight for the GdFeO$_3$ and similar materials.

\begin{acknowledgments}
This work is supported by the Nature Science Foundation of China (No. 11574366), by the Department of Science and Technology of China (Grant No. 2016YFA0300701), and by the Strategic Priority Research Program of the Chinese Academy of Sciences (Grant No.XDB07000000).
\end{acknowledgments}

\end{document}